\pdfoutput=1

\documentclass[iop]{emulateapj}

\slugcomment{}

\shorttitle{Embedded filaments and star formation activities in IRAS 05463+2652}
\shortauthors{L.~K. Dewangan et al.}
\begin{document}

\title{Embedded filaments in IRAS 05463+2652: early stage of fragmentation and star formation activities}
\author{L.~K. Dewangan\altaffilmark{1}, R. Devaraj\altaffilmark{2}, T. Baug\altaffilmark{3} and D.~K. Ojha\altaffilmark{3}}
\email{lokeshd@prl.res.in}
\altaffiltext{1}{Physical Research Laboratory, Navrangpura, Ahmedabad - 380 009, India.}
\altaffiltext{2}{Instituto Nacional de Astrof\'{\i}sica, \'{O}ptica y Electr\'{o}nica, Luis Enrique Erro \# 1, Tonantzintla, Puebla, M\'{e}xico C.P. 72840.}
\altaffiltext{3}{Department of Astronomy and Astrophysics, Tata Institute of Fundamental Research, Homi Bhabha Road, Mumbai 400 005, India.}
\begin{abstract}
We present a multi-wavelength data analysis of IRAS 05463+2652 (hereafter I05463+2652) to study star formation mechanisms.
A shell-like structure around I05463+2652 is evident in the {\it Herschel} column density map, which is not associated with any ionized emission. 
Based on the {\it Herschel} sub-millimeter images, several parsec-scale filaments (including two elongated filaments, ``s-fl" and ``nw-fl" having lengths of $\sim$6.4 and $\sim$8.8 pc, respectively) are investigated in I05463+2652 site. {\it Herschel} temperature map depicts all these features in a temperature range of $\sim$11--13 K.
39 clumps are identified and have masses between $\sim$70--945 M$_{\odot}$. 
A majority of clumps (having M$_{clump}$ $\gtrsim$ 300 M$_{\odot}$) are distributed toward the shell-like structure. 
175 young stellar objects (YSOs) are selected using the photometric 1--5 $\mu$m data and a majority of these YSOs are distributed toward the four areas of high column 
density ($\gtrsim$ 5 $\times$ 10$^{21}$ cm$^{-2}$; A$_{V}$ $\sim$5.3 mag) in the shell-like structure,
where massive clumps and a spatial association with filament(s) are also observed.  
The knowledge of observed masses per unit length of elongated filaments and critical mass length reveals that they are supercritical.
The filament ``nw-fl" is fragmented into five clumps (having M$_{clump}$ $\sim$100--545 M$_{\odot}$) and contains noticeable YSOs, 
while the other filament ``s-fl" is fragmented into two clumps (having M$_{clump}$ $\sim$ 170--215 M$_{\odot}$) without YSOs. 
Together, these observational results favor the role of filaments in star formation process in I05480+2545. 
This study also reveals the filament ``s-fl", containing two starless clumps, at an early stage of fragmentation. 
\end{abstract}
\section{Introduction}
\label{sec:intro}
The investigation of filaments in star-forming regions has got much attention recently, with the availability of the {\it Herschel} continuum data \citep[e.g.,][and references therein]{andre10,andre16,li16,kainulainen16,kainulainen17}. 
There are several theoretical and observational research works available in the literature 
concerning the formation and evolution of filaments \citep[e.g.][and references therein]{ostriker64,inutsuka97,andre10,andre16,li16,kainulainen16,kainulainen17}.
Observationally, it has been reported that the intersections of filaments (or filament mergers) are the potential sites of massive stars and young stellar clusters \citep{schneider12,peretto13,myers09}. 
Furthermore, the filaments are often found to harbor the star-forming clumps and cores along their lengths \citep[e.g.][and references therein]{schneider12,ragan14,contreras16,li16}. 
However, one of the key problems in star formation research is how the filaments fragment into dense clumps/cores that produce star.
It demands careful identification and investigation of the filaments at an early stage of fragmentation and also includes 
a search for starless cores and clumps toward the filaments. 
To our knowledge, such study is still limited in the literature \citep[e.g.,][]{kainulainen16}. 

IRAS 05463+2652 (hereafter I05463+2652) is located at a distance of 2.1 kpc \citep{kawamura98} and is a very poorly explored star-forming site. 
Based on the analysis of the $^{13}$CO (1-0) line data, \citet{kawamura98} examined several molecular clouds in 
Auriga and Gemini including the I05463+2652 site. 
They found an extended molecular cloud associated with I05463+2652 and referred to as ``182.0$-$00.3" cloud (ID \#70; V$_{lsr}$ $\sim$$-$10.6 km s$^{-1}$; line width = 1.7 km s$^{-1}$; radius (R$_{c}$) $\sim$10 pc; Mass (M$_{cloud}$) $\sim$9000 M$_{\odot}$) \citep[see Table 1 and Figure 9j in][]{kawamura98}. 
{\it Herschel} far-infrared (FIR) and sub-millimeter (sub-mm) images are available toward the I05463+2652 site. 
However, these data sets are not yet utilized to infer the embedded filaments, clumps, and physical conditions \citep[e.g.,][]{schneider12,dewangan15,dewangan17a}. 
Hence, the investigation of filaments and their role in star formation processes are yet to be carried out in the I05463+2652 site. 
Furthermore, the study of dust continuum clumps and clusters of young stellar objects (YSOs) in I05463+2652 is still unknown. 
United Kingdom Infra-Red Telescope (UKIRT) Infrared Deep Sky Survey (UKIDSS) near-infrared (NIR) Galactic 
Plane Survey \citep[GPS;][]{lawrence07} photometric data are also available toward I05463+2652 and have better spatial resolution and are deeper than those of Two Micron All Sky Survey data \citep[2MASS;][]{skrutskie06}. 
However, the UKIDSS photometric data sets are yet to be examined in I05463+2652. Hence, a thorough investigation of embedded young stellar populations in the I05463+2652 site is yet to be investigated. 
In this paper, to understand the ongoing star formation processes in the I05463+2652 site, we have performed an 
extensive multi-wavelength study of observations from NIR to radio wavelengths. 
These data sets are collected from numerous surveys 
(such as, the NRAO VLA Sky Survey \citep[NVSS;][]{condon98}, 
the {\it Herschel} Infrared Galactic Plane Survey \citep[Hi-GAL;][]{molinari10}, 
the Wide Field Infrared Survey Explorer \citep[WISE;][]{wright10}, 
the Warm-{\it Spitzer} Galactic Legacy Infrared Mid-Plane Survey Extraordinaire360 \citep[Glimpse360;][]{whitney11}, 
the UKIDSS GPS, and the 2MASS). 
A careful analysis of these data sets enables us to infer the distribution of dust temperature, column density, extinction, ionized emission, 
and YSOs in I05463+2652. 

In Section~\ref{sec:obser}, we present the description of the adopted data sets in this paper. 
Section~\ref{sec:data} presents the results concerning the physical environment and point-like sources.  
Section~\ref{sec:disc} discusses the possible star formation scenario. 
Finally, the main results are summarized in Section~\ref{sec:conc}.
\section{Data and analysis}
\label{sec:obser}
In this work, a target field of $\sim$0$\degr$.825 $\times$ 0$\degr$.825 ($\sim$30.2 pc $\times$ 30.2 pc at a distance of 2.1 kpc) 
(having central coordinates: $l$ = 182$\degr$.224; $b$ = $-$0$\degr$.4) is chosen around I05463+2652. 
In the following, we give a brief description of each data set used in this work. 
\subsection{NIR (1--5 $\mu$m) Data}
In order to assess the embedded young stellar populations in a given star-forming complex, photometric analysis of infrared point-like sources is necessary. 
In this work, to infer embedded YSOs in I05463+2652, deep HK data are examined, along with the {\it Spitzer} Glimpse360 photometry. 
The NIR photometric {\it HK} magnitudes of point-like sources were obtained from the UKIDSS DR10PLUS 
GPS\footnote[1]{http://www.ukidss.org/surveys/gps/gps.html} and the 2MASS\footnote[2]{https://www.ipac.caltech.edu/2mass/}. 
These UKIDSS observations (resolution $\sim$$0\farcs8$) were taken with the Wide Field Camera (WFCAM) mounted 
on the United Kingdom Infra-Red Telescope. The calibration of UKIDSS fluxes was performed using the 2MASS photometric data. 
In this paper, following the selection methods of the GPS photometry 
given in \citet{dewangan15}, we selected only a reliable NIR photometric catalog \citep[see][for more details]{dewangan15}. 
We find the UKIDSS cameras saturated near 11.1 and 10.3 mag in H, and K, respectively.
Hence, the 2MASS photometric magnitudes were obtained for the saturated bright UKIDSS sources. 
To select reliable 2MASS photometric data, we downloaded only those sources having photometric magnitude error of 0.1 or less in each band.

We also utilized ``Warm-{\it Spitzer}" IRAC 3.6 and 4.5 $\mu$m photometric data (resolution $\sim$2$\arcsec$) 
from the Glimpse360\footnote[3]{http://www.astro.wisc.edu/sirtf/glimpse360/} survey. 
The photometric magnitudes of point-like sources were obtained from the highly reliable Glimps360 catalog. 
To increase the significance of our results, only sources with magnitude errors of 0.2 or less were used.
\subsection{Mid-infrared (12--22 $\mu$m) Data}
We retrieved mid-infrared (MIR) images at 12 $\mu$m (spatial resolution $\sim$6$\arcsec$) and 22 $\mu$m (spatial resolution $\sim$1$2\arcsec$) 
from the WISE\footnote[4]{http://irsa.ipac.caltech.edu/Missions/wise.html} database.
\subsection{Far-infrared and Sub-millimeter Data}
We analyzed FIR and sub-mm images retrieved from the {\it Herschel\footnote[5]{http://herschel.cf.ac.uk/mission}} Space Observatory \citep{pilbratt10,poglitsch10,griffin10,degraauw10} data archives. 
Level2$_{-}$5 processed 160--500 $\mu$m images were obtained through the {\it Herschel} Interactive Processing Environment \citep[HIPE,][]{ott10}. 
The beam sizes of the {\it Herschel} images are 12$\arcsec$, 18$\arcsec$, 25$\arcsec$, and 37$\arcsec$ for 160, 250, 350, and 500 $\mu$m, respectively \citep{poglitsch10,griffin10}. 
The plate scales of 160, 250, 350, and 500 $\mu$m images are 3$''$.2, 6$''$, 10$''$, and 14$''$ pixel$^{-1}$, respectively.  
The {\it Herschel} images at 250--500 $\mu$m are calibrated in units of surface brightness, MJy sr$^{-1}$, while the unit of image at 160 $\mu$m 
is Jy pixel$^{-1}$. 

Using the {\it Herschel} 160--500 $\mu$m images, the {\it Herschel} temperature and column density maps of I05463+2652 have been produced. 
Following the procedures mentioned in \citet{mallick15}, these maps are obtained from a 
pixel-by-pixel spectral energy distribution (SED) fit with a modified blackbody to the cold dust emission at {\it Herschel} 160--500 $\mu$m \citep[also see][]{dewangan15}. 
In the following, we give a brief step-by-step explanation of the adopted methods. 

Prior to the SED fit, using the task ``Convert Image Unit" available in the HIPE software, the {\it Herschel} images at 250--500 $\mu$m were 
converted into the flux unit (i.e. Jy pixel$^{-1}$) of the image at 160 $\mu$m. 
Using the plug-in ``Photometric Convolution" available in the HIPE software, the 160--350 $\mu$m images were transferred 
to a common grid with the same resolution and pixel size of 37$\arcsec$ and 14$\arcsec$, respectively.
These values (resolution and pixel size) are the parameters of the 500 $\mu$m image.
Next, the sky background flux level was computed to be 0.060, 0.133, 0.198, and $-$0.095 Jy pixel$^{-1}$ for the 500, 350, 250, and 
160 $\mu$m images (size of the selected featureless dark region $\sim$13$\farcm$4 $\times$ 14$\farcm$8; 
centered at:  $l$ = 183$\degr$.181; $b$ = $-$0$\degr$.354), respectively. 

In the final step, to generate the temperature and column density maps, a modified blackbody was fitted to the observed fluxes on a pixel-by-pixel basis 
\citep[see equations 8 and 9 in][]{mallick15}. 
The fitting was done using the four data points for each pixel, keeping the column density ($N(\mathrm H_2)$) and the 
dust temperature (T$_{d}$) as free parameters. 
In the analysis, we considered a mean molecular weight per hydrogen molecule ($\mu_{H2}$) of 2.8 
\citep{kauffmann08} and an absorption coefficient ($\kappa_\nu$) of 0.1~$(\nu/1000~{\rm GHz})^{\beta}$ cm$^{2}$ g$^{-1}$, 
including a gas-to-dust ratio ($R_t$) of 100, with a dust spectral index ($\beta$) of 2 \citep[see][]{hildebrand83}. 
The resultant temperature and column density maps (resolution $\sim$37$\arcsec$) are discussed in Section~\ref{subsec:temp}.
\subsection{Radio continuum data}
The radio continuum map at 1.4 GHz (21 cm; beam size  $\sim$45$\arcsec$) was obtained from the NVSS archive.
The radio continuum emission traces the ionized emission.
\section{Results}
\label{sec:data}
\subsection{Physical environment of IRAS 05463+2652}
\label{subsec:u1}
In this section, we present {\it WISE}, {\it Herschel}, and NVSS images of I05463+2652 to examine its morphology, and 
the distribution of cold dust and ionized emissions.

\begin{figure*}
\epsscale{0.72}
\plotone{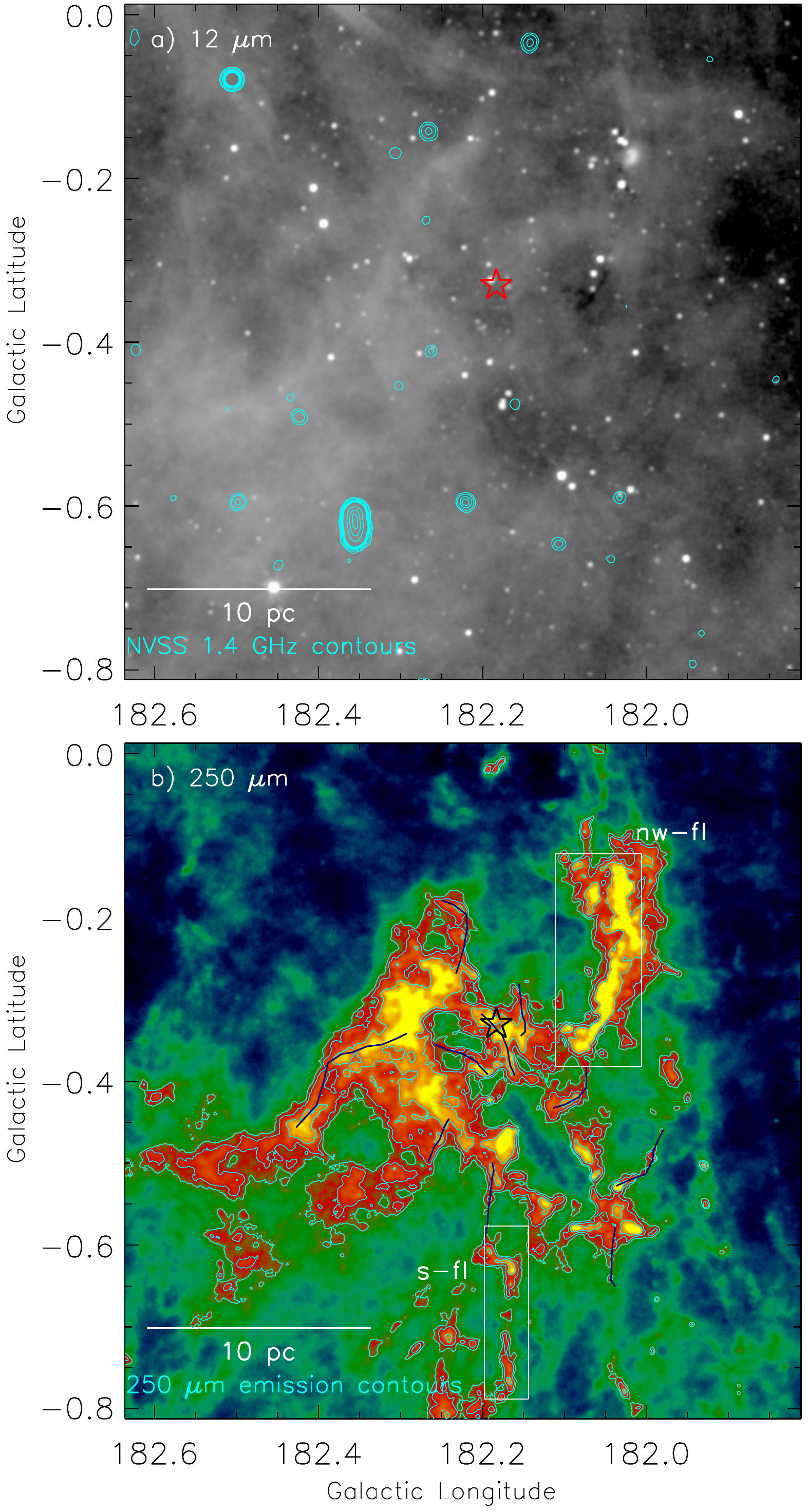}
\caption{\scriptsize {\it WISE} 12 $\mu$m and {\it Herschel} 250 $\mu$m images of I05463+2652 (size of the chosen field $\sim$0$\degr$.825 $\times$ 0$\degr$.825 
($\sim$30.2 pc $\times$ 30.2 pc at a distance of 2.1 kpc); 
central coordinates: $l$ = 182$\degr$.224; $b$ = $-$0$\degr$.4). 
a) A gray-scale 12 $\mu$m image (in log scale) is superimposed with the NVSS 1.4 GHz contours. 
The NVSS 1.4 GHz contours (in cyan) are shown with levels of 0.55, 1, 2, 3, 4, 5, 6, 30, 60, and 90\% of the peak value (i.e.  0.377 Jy/beam).
b) A false color 250 $\mu$m continuum image is overlaid with the 250 $\mu$m emission contours. 
The contours (in cyan) are shown with levels of 0.078, 0.085, and 0.098 Jy/pix. 
Several filaments are highlighted by curves (in navy). The boxes also encompass the elongated filamentary structures in the image (i.e. ``nw-fl" and ``s-fl"; see text for details). 
In both the panels, the position of I05463+2652 is highlighted by a star.
In each panel, the scale bar corresponding to 10 pc (at a distance of 2.1 kpc) is shown in the bottom left corner.}
\label{fig1}
\end{figure*}

\begin{figure*}
\epsscale{0.72}
\plotone{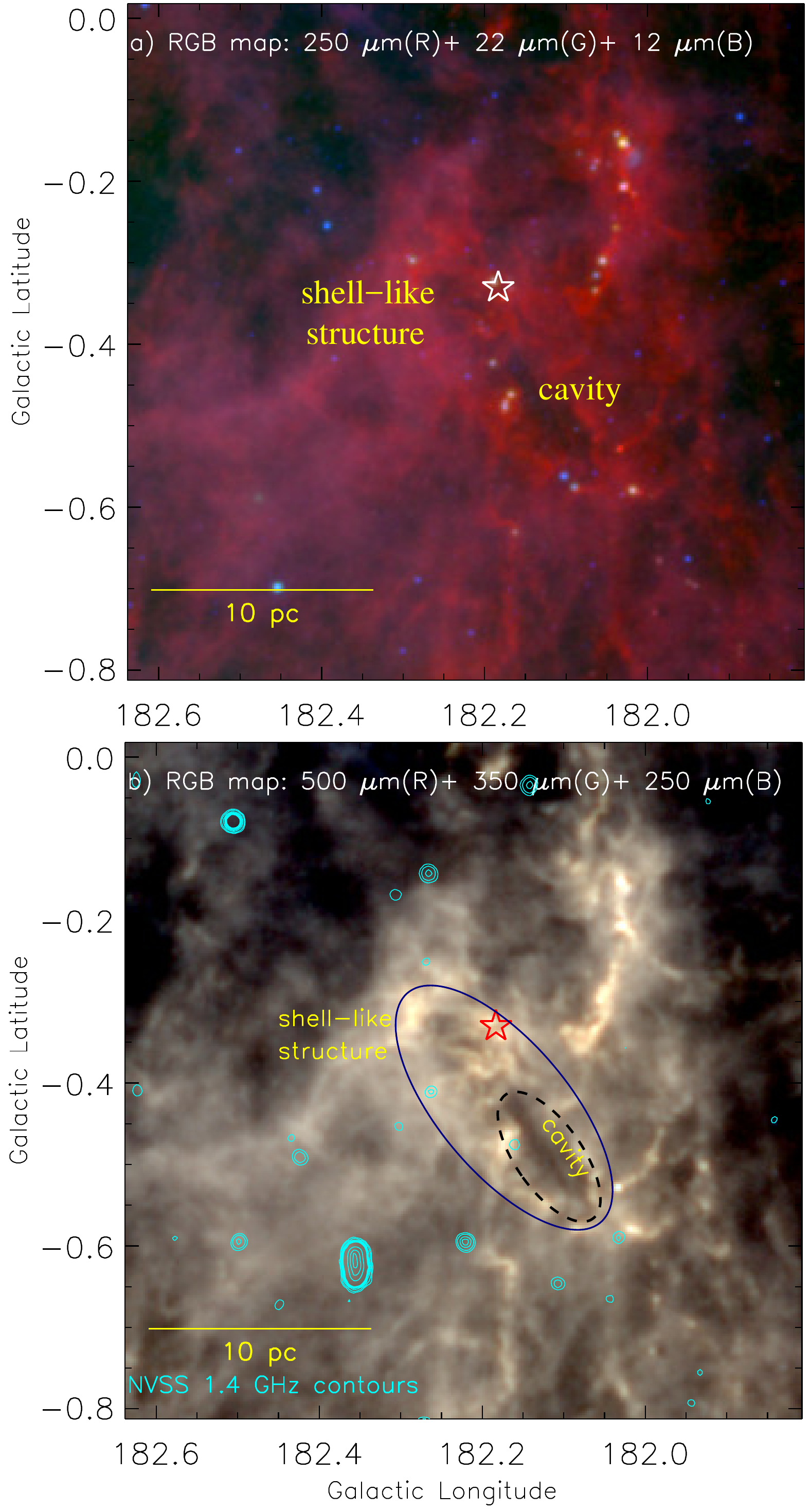}
\caption{\scriptsize Multi-wavelength view of I05463+2652 . 
a) A three color-composite map ({\it Herschel} 250 $\mu$m (red), {\it WISE} 22 $\mu$m (green), and {\it WISE} 12 $\mu$m (blue) images 
in log scale). b) A three color-composite map ({\it Herschel} 500 $\mu$m (red), 350 $\mu$m (green), and 250 $\mu$m (blue) images 
in log scale). Using two ellipses, a shell-like structure and a cavity are also highlighted. 
The composite map is also overlaid with the NVSS 1.4 GHz contours (in cyan) which are similar to those shown in Figure~\ref{fig1}a. In each panel, a shell-like structure and a cavity are labeled. 
In both the panels, the position of I05463+2652 is highlighted by a star.
In each panel, the scale bar corresponding to 10 pc (at a distance of 2.1 kpc) is shown in the bottom left corner.}
\label{fig2}
\end{figure*}

Figures~\ref{fig1}a and~\ref{fig1}b show the {\it WISE} 12 $\mu$m and {\it Herschel} 250 $\mu$m images of I05463+2652, respectively.
The NVSS 1.4 GHz emission is also overlaid on the {\it WISE} 12 $\mu$m image (see Figure~\ref{fig1}a). 
No ionized emission is observed toward the IRAS position.
In our selected field, the 12 $\mu$m image also shows the presence of embedded stellar contents and some of these may be infrared excess sources (see Section~\ref{subsec:phot1}). 
The {\it Herschel} 250 $\mu$m image is also superimposed with the 250 $\mu$m emission contours, indicating the presence of extended features in our selected site. 
The {\it Herschel} image also enables us to depict several embedded filaments in our selected site around I05463+2652 (see Figure~\ref{fig1}b). These filaments are identified by visual inspection and are shown by curves in the figure. 
We have also highlighted two other elongated filamentary features in the 250 $\mu$m image. 
The filament seen in the galactic north-west direction is referred to as ``nw-fl" (length $\sim$8.8 pc),
while the other one detected in the galactic south direction is designated to as ``s-fl" (length $\sim$6.4 pc) (see boxes and labels in Figure~\ref{fig1}b and also Section~\ref{subsec:temp} for more details).
In Figure~\ref{fig2}a, we present a three-color composite map made using the {\it Herschel} 250 $\mu$m (in red), {\it WISE} 22 $\mu$m (in green), and 
{\it WISE} 12 $\mu$m (in blue) images. We find noticeable embedded stellar sources toward some filaments and have presented quantitative analysis of the infrared excess sources in Section~\ref{subsec:phot1}. 
Figure~\ref{fig2}b shows a color-composite map produced using the {\it Herschel} sub-mm images 
(i.e. 500 $\mu$m (red), 350 $\mu$m (green), and 250 $\mu$m (blue)). The map is also overlaid with the NVSS 1.4 GHz continuum emission. 
Interestingly, an embedded shell-like structure is seen in the {\it Herschel} images (see a big ellipse in Figure~\ref{fig2}b). 
Additionally, there is a central dark region toward the shell-like structure (see a dashed ellipse in 
Figure~\ref{fig2}b), which appears to be a cavity without any condensations and ionized emission.  
Hence, it is unlikely that this structure is originated by the feedback of any massive stars.
In our selected field, the {\it Herschel} images also reveal several condensations toward the shell-like structure and filaments 
(see Section~\ref{subsec:temp} for quantitative analysis).
It is also observed that a majority of selected filaments seem to be connected to the shell-like structure (see Figures~\ref{fig1}b and~\ref{fig2}b), while the two elongated filamentary features (i.e. ``nw-fl" and ``s-fl") appear far more spatially isolated from the shell-like feature.

To compute virial mass of the molecular cloud linked with I05463+2652 \citep[i.e., ``182.0$-$00.3";][]{kawamura98}, 
we have utilized the $^{13}$CO line derived parameters (e.g., line-width ($\Delta V$ =1.7 km s$^{-1}$), radius (R$_{c}$=10 pc), 
and mass (M$_{cloud}$) = 9000 M$_{\odot}$). 
The virial mass (M$_{vir}$) of a cloud of radius R$_{c}$ (in pc) and line-width $\Delta V$ (in km s$^{-1}$) is defined 
as M$_{vir}$ ($M_\odot$)\,=\,k\,R$_{c}$\,$\Delta V^2$ \citep{maclaren88}, 
where the geometrical parameter, k\,=\,126, for a density profile $\rho$ $\propto$ 1/r$^2$. 
A virial parameter is the ratio of the virial mass of a cloud to its actual mass (i.e. M$_{vir}$/M$_{cloud}$), 
helping to infer stability against collapse. The virial parameter less than 1 hints the cloud prone to collapse, and greater than 1 is against the collapse. 
We obtain M$_{vir}$ $\sim$3640 $M_\odot$, which is less than M$_{cloud}$ (i.e. M$_{vir}$ $<$ M$_{cloud}$). 
This implies that the cloud is unstable against gravitational collapse, explaining the existence of several condensations in the I05463+2652 site.

Together, Figures~\ref{fig1} and~\ref{fig2} allow us to probe the embedded shell-like appearance of I05463+2652, 
several embedded filaments, and condensations in our selected site. 
\subsection{{\it Herschel} temperature and column density maps}
\label{subsec:temp}
\begin{figure*}
\epsscale{0.7}
\plotone{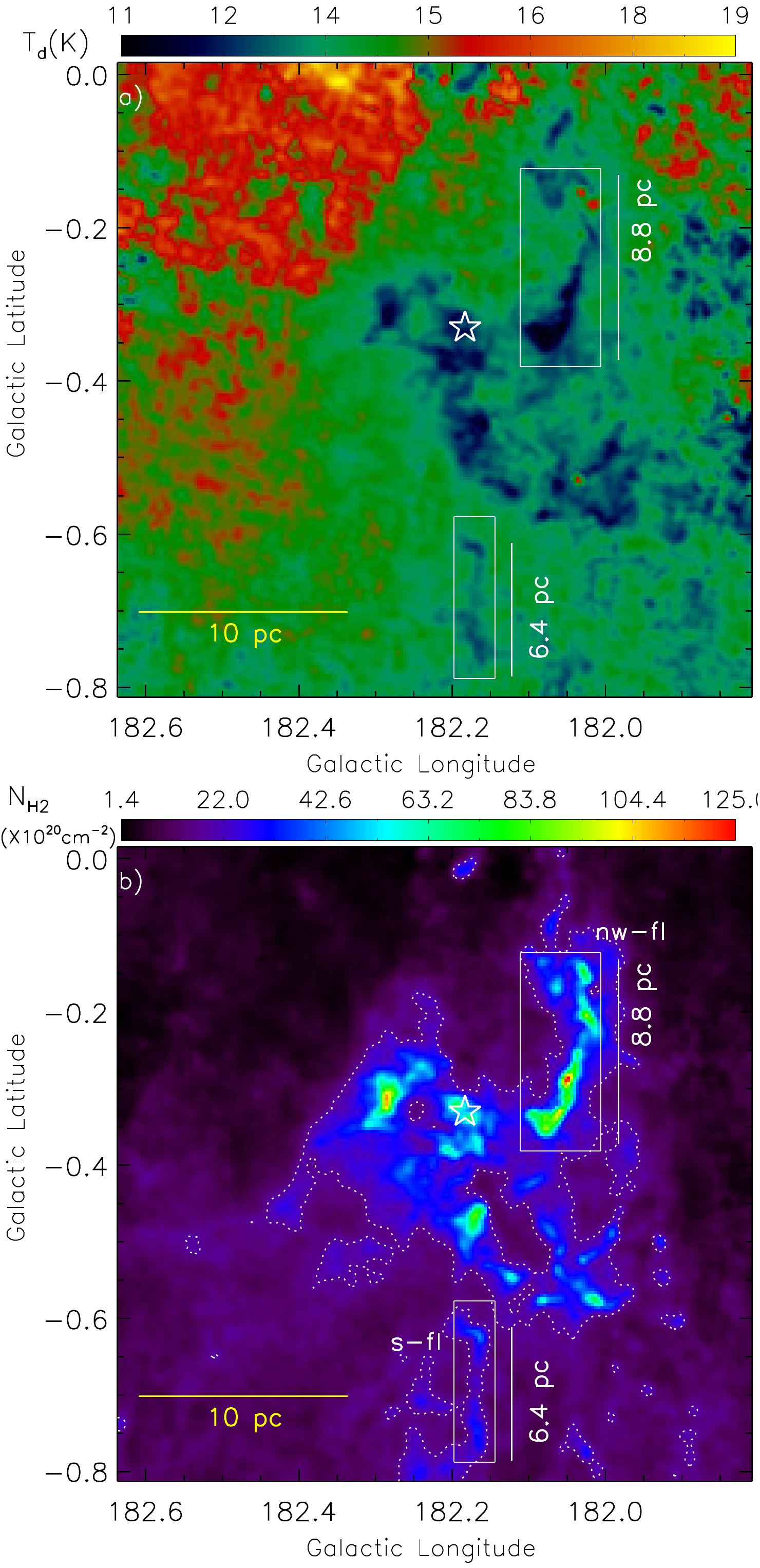}
\caption{\scriptsize a) {\it Herschel} temperature map of I05463+2652. 
b) {\it Herschel} column density ($N(\mathrm H_2)$) map of I05463+2652. 
The column density map also enables to obtain the extinction using the relation $A_V=1.07 \times 10^{-21}~N(\mathrm H_2)$ \citep{bohlin78}. 
The map is also overlaid with a column density contour (in dotted while) and the contour level is 1.8 $\times$ 10$^{21}$ cm$^{-2}$ (A$_{V}$ $\sim$2 mag). 
In both the panels, the boxes encompass the elongated filamentary structures (``nw-fl" and ``s-fl") and the position of I05463+2652 is highlighted by a star.
In each panel, the scale bar corresponding to 10 pc (at a distance of 2.1 kpc) is shown in the bottom left corner.}
\label{fig3}
\end{figure*}

\begin{figure*}
\epsscale{1}
\plotone{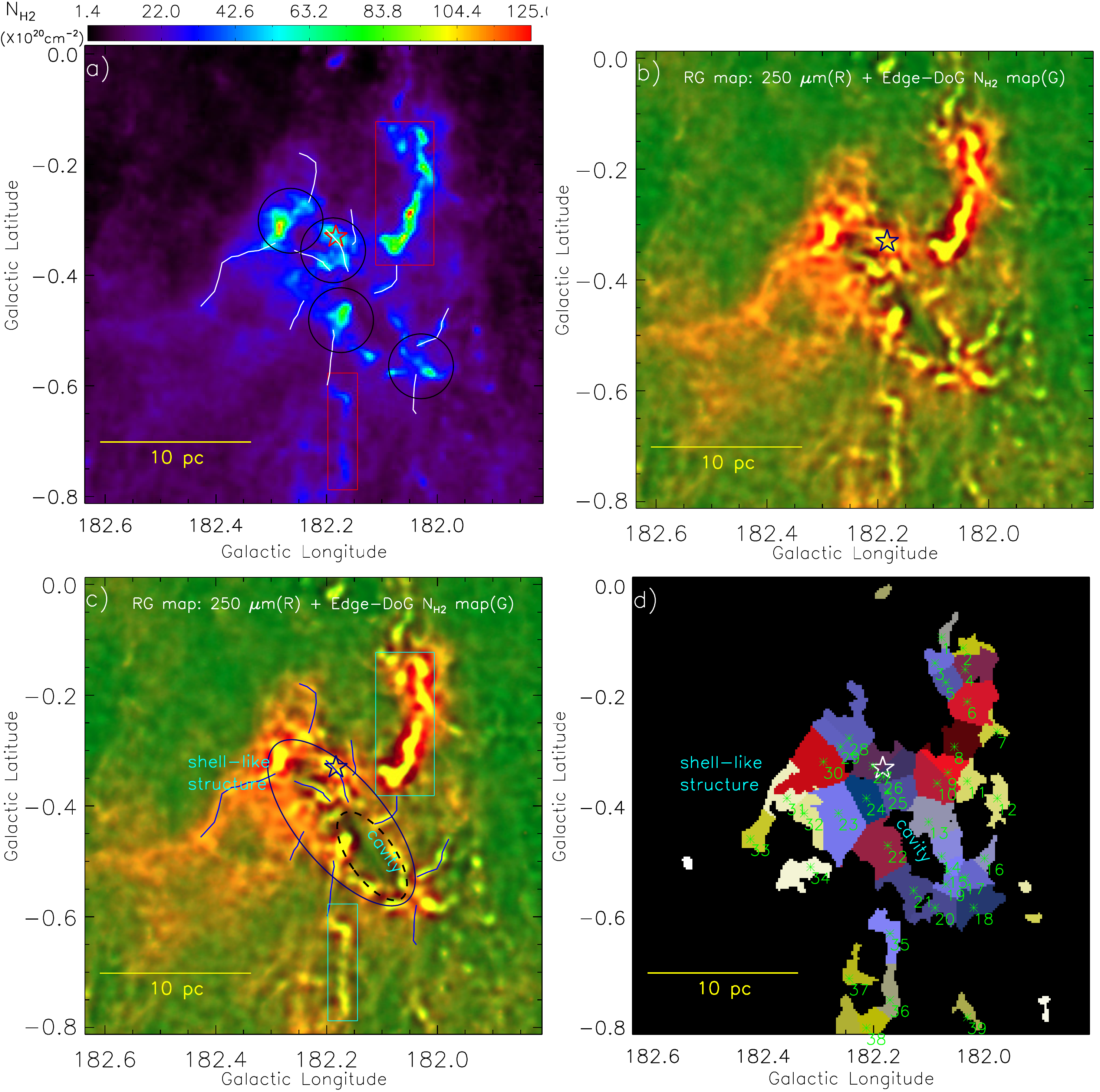}
\caption{\scriptsize a) {\it Herschel} column density ($N(\mathrm H_2)$) map of I05463+2652. 
Several filaments are highlighted by curves which are similar to those shown in Figure~\ref{fig1}b. 
Black circles show at least four areas of high column density ($\gtrsim$ 5 $\times$ 10$^{21}$ cm$^{-2}$ (A$_{V}$ $\sim$5.3 mag)). 
The boxes also encompass two other elongated filamentary structures (``nw-fl" and ``s-fl"; also see Figure~\ref{fig3}b). 
b) A two color-composite map ({\it Herschel} 250 $\mu$m (red) and {\it Herschel} column density map (green) images) of I05463+2652. 
Here, the column density map is processed through an ``Edge-DoG" algorithm. 
c) It is the same as shown in Figure~\ref{fig3x}b. A shell-like structure and a cavity are labeled and are also highlighted by ellipses (also see Figure~\ref{fig2}b). Other marked curves and boxes are similar to those shown in Figure~\ref{fig3x}a. 
d) The distribution of identified {\it Herschel} clumps in our probed field around I05463+2652.
The identified clumps are highlighted by asterisks and the boundary of each clump is also shown in 
the figure along with its corresponding clump ID (see Table~\ref{tab1}). A shell-like structure and a cavity are also labeled.
In all the panels, the position of I05463+2652 is highlighted by a star.
In each panel, the scale bar corresponding to 10 pc (at a distance of 2.1 kpc) is shown in the bottom left corner.}
\label{fig3x}
\end{figure*}

In this section, to further investigate the shell-like structure, filaments, and condensations, 
the {\it Herschel} temperature and column density maps of I05463+2652 are examined. 
The final temperature and column density maps (resolution $\sim$37$\arcsec$) are shown in Figures~\ref{fig3}a and~\ref{fig3}b, respectively.
The temperature map reveals the shell-like structure and filaments in a temperature range of about 11--13~K.
The column density map also traces the shell-like structure, filaments (including two elongated filamentary features, ``nw-fl" and ``s-fl"), and condensations, allowing quantitative analysis of these features.  
In the column density map, the elongated filamentary features are depicted by a column density contour level of 1.8 $\times$ 10$^{21}$ cm$^{-2}$ (corresponding A$_{V}$ $\sim$2 mag).
Here, we have used a relation between optical extinction and hydrogen column density 
\citep[i.e. $A_V=1.07 \times 10^{-21}~N(\mathrm H_2)$;][]{bohlin78}. 
The noticeable condensations are also seen toward these two elongated filamentary features.
Several other filaments are also highlighted by curves in the column density map (see also Figure~\ref{fig3x}a and Figure~\ref{fig1}b). As mentioned before, these filaments are identified based on the visual inspection. 
In Figure~\ref{fig3x}a, black circles indicate at least four areas of high column density 
($\gtrsim$ 5 $\times$ 10$^{21}$ cm$^{-2}$; corresponding A$_{V}$ $\sim$5.3 mag), 
where these filaments are observed. 
To further visually reveal the embedded filaments, Figure~\ref{fig3x}b shows a two color-composite image produced using the {\it Herschel} 
250 $\mu$m (red) and {\it Herschel} column density map (green) images. 
An edge detection algorithm \citep[i.e. Difference of Gaussian (DoG); see][]{gonzalez11,assirati14} has been applied to the {\it Herschel} 
column density map. In this technique, two Gaussian kernels are subtracted, where a kernel has a standard deviation relatively smaller than the previous one \citep[i.e.][]{assirati14}. We employed two radius values (i.e.  3 and 5 pixels) of the Gaussians in this work. 
The structures revealed by the DoG algorithm depend on the difference between these two radius values.
If one uses larger radius values ($>$ 6 pixels) then the features seen in the resultant map become more blurry. 
The DoG processed column density map enables us to obtain better visual appearance of the embedded features in our selected target field.  
In Figure~\ref{fig3x}b, the filaments as well as the shell-like structure are clearly seen in the color-composite map. 
 Using the same color-composite map, 
we have also highlighted the shell-like structure, cavity, and 
filaments in Figure~\ref{fig3x}c. 

In the column density map, we have employed the ``{\it clumpfind}" IDL program \citep{williams94} to identify the clumps and 
to compute their total column densities. 
39 clumps are found in the map and are labeled in Figure~\ref{fig3x}d. 
Furthermore, the boundary of each clump is also shown in Figure~\ref{fig3x}d. 
We have also computed the mass of each clump using its total column density. 
The mass of a single {\it Herschel} clump is estimated using the following formula:
\begin{equation}
M_{clump} = \mu_{H_2} m_H A_{pix} \Sigma N(H_2)
\end{equation}
where $\mu_{H_2}$ is assumed to be 2.8, $A_{pix}$ is the area subtended by one pixel, and 
$\Sigma N(\mathrm H_2)$ is the total column density. 
The mass of each {\it Herschel} clump is listed in Table~\ref{tab1}. 
The table also gives an effective radius of each clump, which is an outcome of the {\it clumpfind} algorithm. 
It is obvious from Table~\ref{tab1} that the clump masses vary between 70 M$_{\odot}$ and 945 M$_{\odot}$. 
A majority of clumps (having M$_{clump}$ $\gtrsim$ 300 M$_{\odot}$) are distributed toward the shell-like feature 
(see clump nos 13 to 32 in Figure~\ref{fig3x}d and also Table~\ref{tab1}). 
Figure~\ref{fig3x}d also helps us to infer the embedded shell-like structure and the cavity. 
Note that, at least four areas of high column density (see black circles in Figure~\ref{fig3x}a) 
are observed, where the massive clumps are also detected. 
In Figure~\ref{fig3x}a, we also find at least two filaments traced in the three areas of high column density highlighted by circles.

Concerning to the elongated filament ``nw-fl" (length $\sim$8.8 pc), we have found five clumps (nos 2, 4, 6, 8, and 9) 
that have masses between $\sim$100--545 M$_{\odot}$ and the total mass of these clumps is 1760 M$_{\odot}$ (see Figure~\ref{fig3x}d and also Table~\ref{tab1}). 
Using the mass and length of the filamentary feature, the mass per unit length is estimated to be $\sim$200 M$_{\odot}$ pc$^{-1}$. 
Furthermore, there are at least two clumps (nos 35 (M$_{clump}$$\sim$215 M$_{\odot}$) and 36 (M$_{clump}$$\sim$170 M$_{\odot}$)) 
seen toward the elongated filament ``s-fl" (length $\sim$6.4 pc), 
allowing to compute the mass per unit length of $\sim$60 M$_{\odot}$ pc$^{-1}$. 
We do not know the inclination angle, i, of each elongated filament, and for reference, we have assumed here i = 0.
Due to the inclination, the line mass can be affected by a factor of cos i \citep[e.g.,][]{kainulainen16}. 
Hence, the observed mass per unit length values can be taken as an upper limit. 
Note that a critical line mass ($M_{\rm line,crit}~=~2 c^{2}_{\rm s}/G$; where $c_{\rm s}$ is the isothermal sound speed and $G$ is the gravitational constant) 
is reported to be $\sim$16~M$_{\odot}$ pc$^{-1}$ for gas filaments at T= 10~K 
(where $c_{\rm s} \sim 0.2$~km s$^{-1}$) 
\citep[e.g.,][]{ostriker64,andre14,kainulainen16}. 
Hence, the observed masses per unit length of elongated filamentary features are greater than the critical mass per unit length at T = 10 K. 
Note that even if we assume 10--50\% uncertainty in the estimation of mass per unit length for each elongated filament (due to inclination and distance uncertainties), the observed mass per unit length for each elongated filament is still higher than the critical value.
Together, the analysis also indicates that these two filamentary features appear to be gravitationally unstable to 
radial contraction and fragmentation along its length \citep[e.g.][]{inutsuka97}. 
\subsection{Young stellar populations in the IRAS 05463+2652 site}
\label{subsec:phot1}
\subsection{Selection of YSOs}
In this section, we have investigated the embedded young stellar populations with their infrared-excess emission in the I05463+2652 site.

\begin{figure*}
\epsscale{0.92}
\plotone{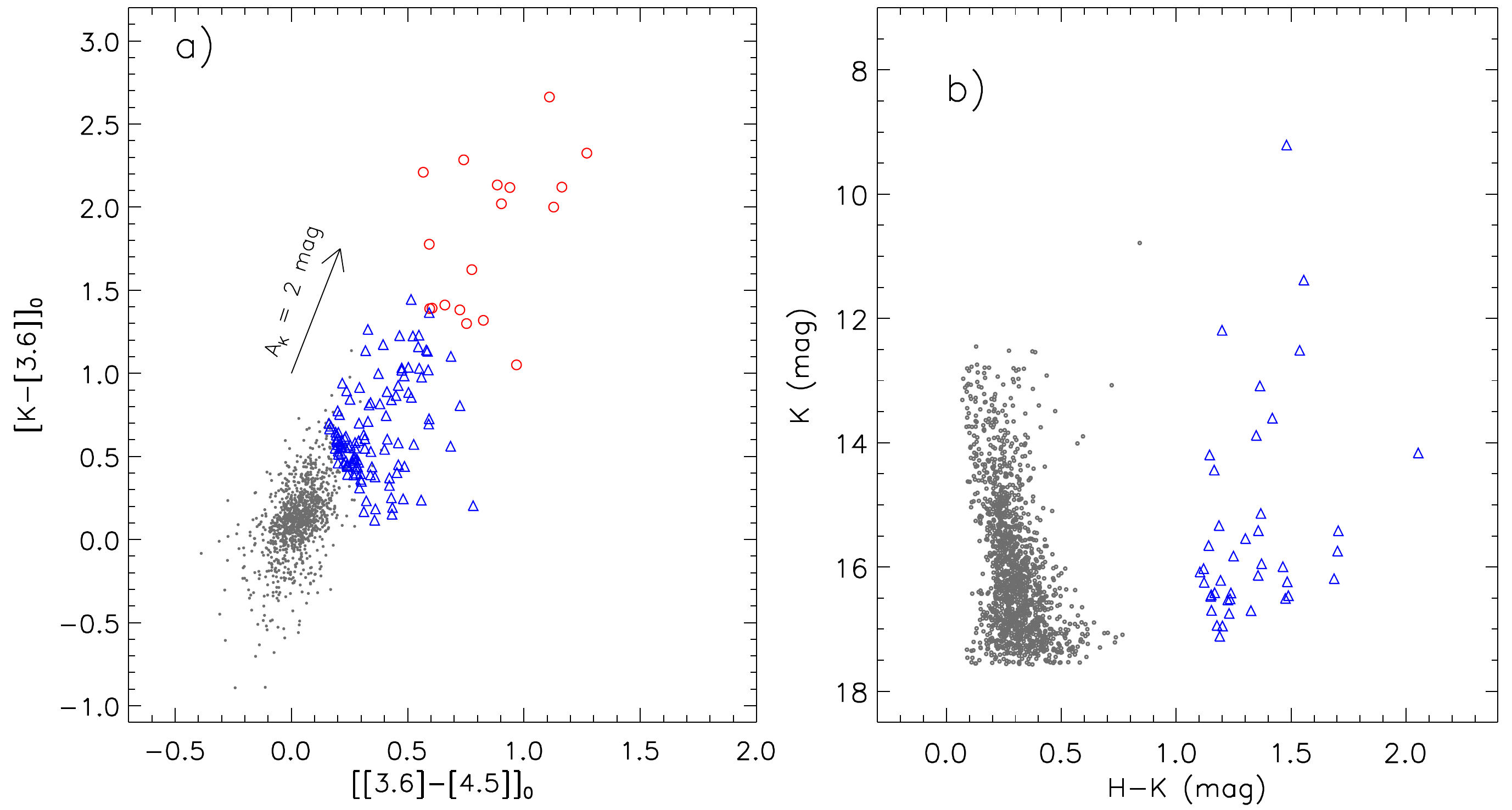}
\epsscale{0.76}
\plotone{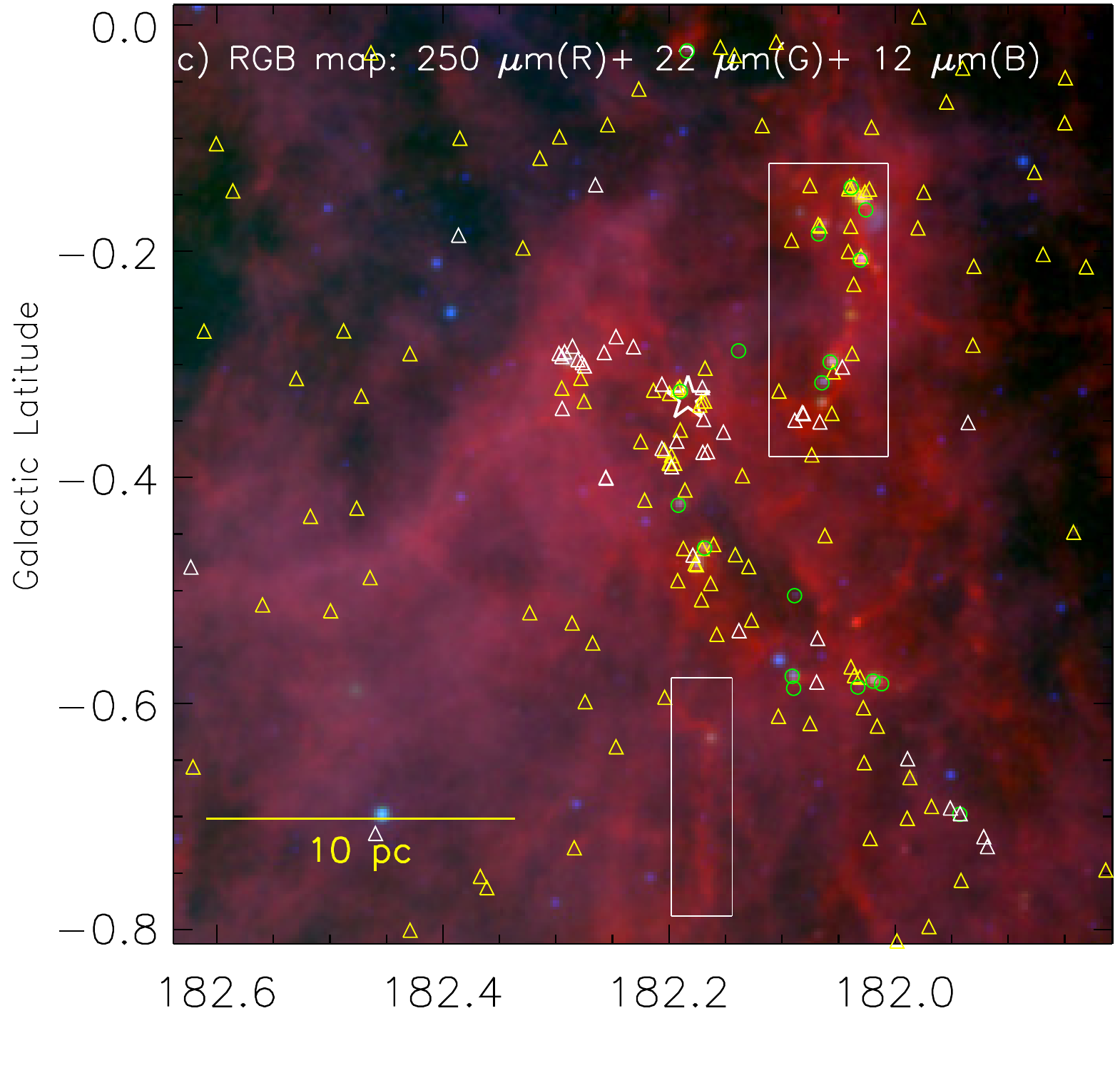}
\caption{\scriptsize The selection of young stellar populations in our probed field around I05463+2652. 
a) The panel shows a dereddened color-color diagram ([K$-$[3.6]]$_{0}$ $vs$ [[3.6]$-$[4.5]]$_{0}$) constructed 
using the H, K, 3.6 $\mu$m, and 4.5 $\mu$m data (see text for details). 
The extinction vector is shown in the panel and is obtained using the average extinction laws from \citet{flaherty07}. 
b) The panel shows a color-magnitude diagram (H$-$K/K) of sources detected only in H and K bands that have no counterparts 
in our selected GLIMPSE360 catalog. c) The spatial distribution of selected YSOs. The positions of YSOs are overlaid on a color-composite map, 
which is similar to the one shown in Figure~\ref{fig2}a. 
The boxes encompass two elongated filamentary structures (``nw-fl" and ``s-fl"; also see Figure~\ref{fig3}b). 
The position of I05463+2652 is highlighted by a star.
The scale bar corresponding to 10 pc (at a distance of 2.1 kpc) is shown in the bottom left corner. 
In the first two panels, circles and open triangles refer to Class~I and Class~II YSOs, respectively, and 
the dots (in gray) indicate the stars with only photospheric emission. 
In the color-color diagram, we have shown only 1101 out of 8520 stars with photospheric emission. 
In the color-magnitude diagram, we have shown only 1501 out of 21662 stars with photospheric emissions. 
Due to large numbers of stars with photospheric emissions, we have randomly shown only some of these stars in the diagrams. 
In the last panel, the YSOs selected using the H, K, 3.6 $\mu$m, and 4.5 $\mu$m data 
are shown by circles (in green) and triangles (in yellow) (see Figure~\ref{fig4}a), whereas the white triangles represent the YSOs 
identified using the H and K bands (see Figure~\ref{fig4}b).} 
\label{fig4} 
\end{figure*}

The dereddened color-color diagram ([K$-$[3.6]]$_{0}$ and [[3.6]$-$[4.5]]$_{0}$) is a promising tool to find infrared-excess sources \citep[e.g.][]{gutermuth09}. 
Using the 2MASS and GLIMPSE360 photometric data at 1--5 $\mu$m, we have examined the dereddened color-color space ([K$-$[3.6]]$_{0}$ and [[3.6]$-$[4.5]]$_{0}$). The dereddened colors were estimated using the color excess ratios given in \citet{flaherty07}. 
Following the dereddened color conditions listed in \citet{gutermuth09}, we find 134 (18 Class~I and 116 Class~II) YSOs in our selected field. 
One can also remove possible dim extragalactic contaminants from the identified YSOs with the help of previously known conditions (i.e., [3.6]$_{0}$ $<$ 15 mag for Class~I and [3.6]$_{0}$ $<$ 14.5 mag for Class~II) \citep[e.g.,][]{gutermuth09}.  
The dereddened 3.6 $\mu$m magnitudes were computed using the observed color and the reddening laws \citep[from][]{flaherty07}.
Figure~\ref{fig4}a shows the color-color diagram ([K$-$[3.6]]$_{0}$ versus [[3.6]$-$[4.5]]$_{0}$).  
The red circles and blue triangles refer to Class~I and Class~II YSOs, respectively.\\  
Furthermore, using the 2MASS and UKIDSS-GPS data, we have also made a color-magnitude diagram (H$-$K/K) to obtain additional YSOs in the I05463+2652 site 
(see Figure~\ref{fig4}b). Note that we have chosen sources observed only in H and K bands that have no counterparts in our selected GLIMPSE360 catalog. 
The diagram indicates the embedded sources having H$-$K $>$ 1.1 mag. 
This color criterion was chosen based on the color-magnitude analysis of a nearby control field (selected size $\sim$0$\degr$.2 
$\times$ 0$\degr$.2; central coordinates: $l$ = 181$\degr$.697; $b$ = $-$0$\degr$.487). 
This particular scheme gives 41 additional YSOs in the selected field.\\ 

Together, these two schemes yield a total of 175 YSOs in the selected field. 
In Figure~\ref{fig4}c, all the identified YSOs are overlaid on a color-composite map made using the {\it Herschel} and {\it WISE} images.
Furthermore, we have also overplotted the YSOs on the {\it Herschel} column density map (see Figure~\ref{fig5}a), tracing 
a majority of YSOs toward the areas of high column density. 
Additionally, noticeable YSOs are also observed toward the elongated filament ``nw-fl".
\subsubsection{Clustering of YSOs}
\label{subsec:srf}
\begin{figure*}
\epsscale{0.72}
\plotone{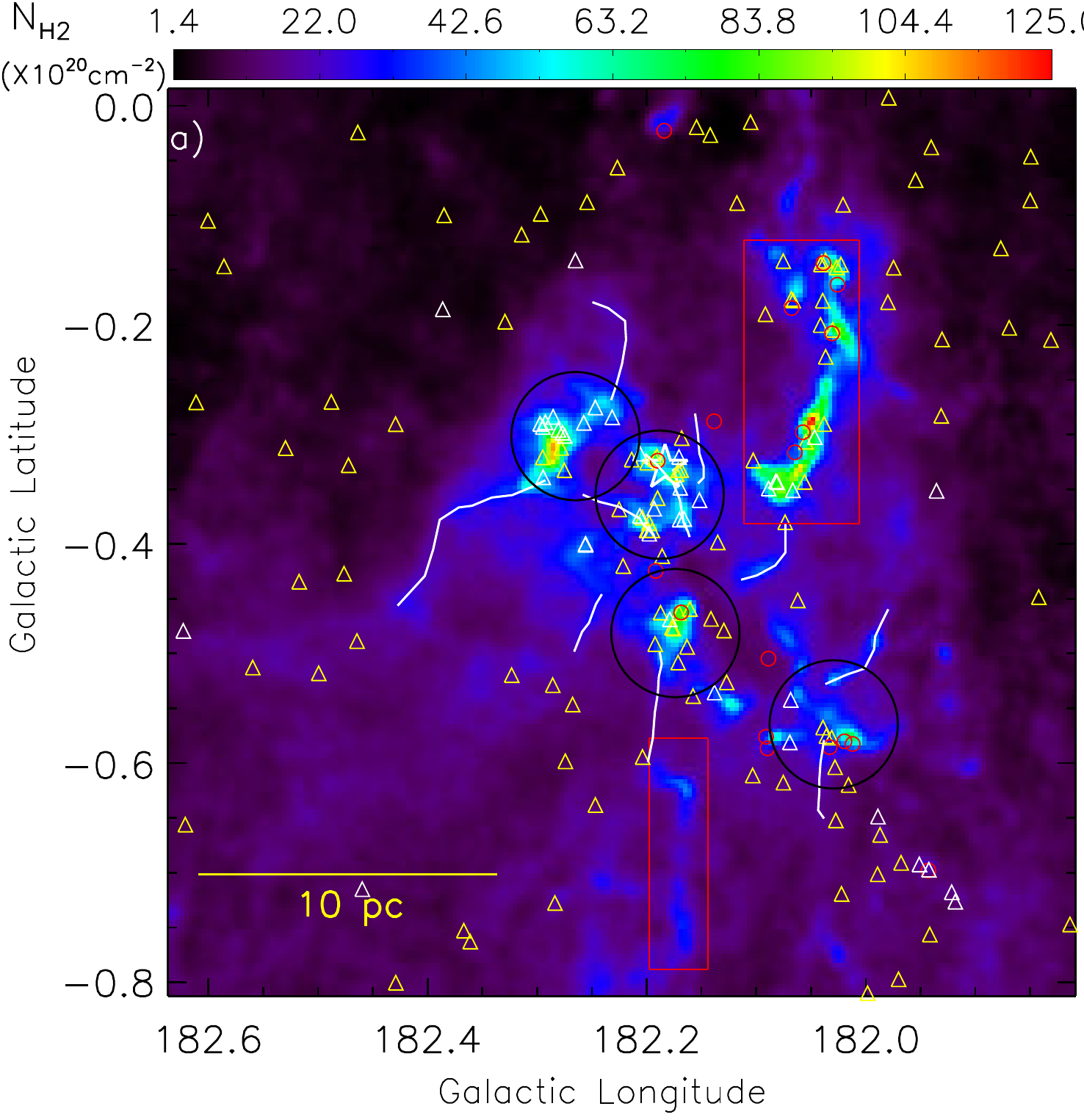}
\epsscale{0.72}
\plotone{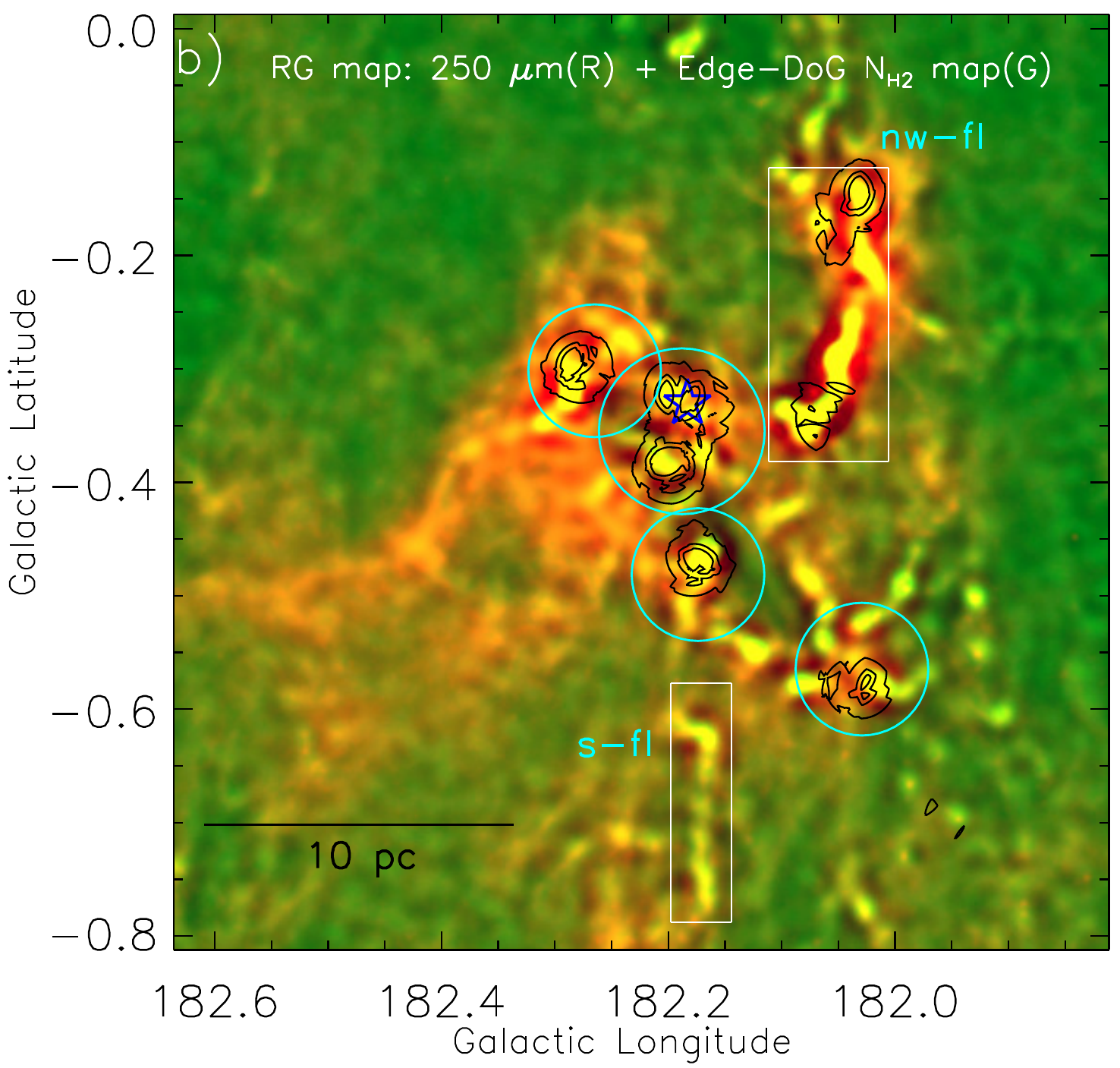}
\caption{\scriptsize a) {\it Herschel} column density map is overlaid with the identified YSOs in our selected field around I05463+2652.
The marked symbols are similar to those shown in Figure~\ref{fig4}c. Several filaments are highlighted by curves which are similar to 
those shown in Figures~\ref{fig1}b and~\ref{fig3x}a. Black circles indicate the areas of high column 
density ($\gtrsim$ 5 $\times$ 10$^{21}$ cm$^{-2}$ (A$_{V}$ $\sim$5.3 mag)). The background map is similar to the one shown in Figure~\ref{fig3}b. 
b) A two color-composite map ({\it Herschel} 250 $\mu$m (red) and {\it Herschel} column density map (green) images) 
is superimposed with the surface density contours (in black) of YSOs. 
The background map is similar to the one shown in Figure~\ref{fig3x}b. 
The maps show the presence of the clusters of YSOs toward the areas of high column 
density in the shell-like structure and the filament ``nw-fl". 
The contours are shown at [3, 6, 9] $\times$ $\sigma$ (where, 1$\sigma$ = 0.4 YSOs/pc$^{2}$), from the 
outer to the inner side. In both the panels, the boxes encompass two elongated filamentary structures (``nw-fl" and ``s-fl"; also see Figure~\ref{fig3}b). 
In each panel, the position of I05463+2652 is highlighted by a star and 
the scale bar corresponding to 10 pc (at a distance of 2.1 kpc) is shown in the bottom left corner.} 
\label{fig5} 
\end{figure*}

In this section, we have examined the presence of clustering of YSOs in I05463+2652 based on their surface density map. 
The surface density map of YSOs is produced using the nearest-neighbour (NN) method \citep[see][for more details]{gutermuth09,bressert10}. 
The surface density map of 175 YSOs is generated, in a manner similar to that utilized in \citet{dewangan15}. 
The map is obtained using a 5$\arcsec$ grid and 6 NN at a distance of 2.1 kpc. 
Figure~\ref{fig5}b shows the resultant surface density contours of YSOs superimposed on a two color-composite image, 
which is similar to the one shown in Figure~\ref{fig3x}b. 
The contours are shown at [3, 6, 9] $\times$ $\sigma$ (where, 1$\sigma$ = 0.4 YSOs/pc$^{2}$), from the 
outer to the inner regions. 
The low values of the observed surface density do not support the presence of intense star formation activities in the I05463+2652 site.
However, the clusters/groups of YSOs are found toward the four areas of high column density in the shell-like structure (see black circles in Figure~\ref{fig5}b), where the massive clumps are also found (also see Figure~\ref{fig5}a).  
Furthermore, the clusters of YSOs are also seen toward the elongated filament ``nw-fl".
But, the elongated filament ``s-fl" is not associated with any YSOs and their clustering.

Together, the star formation activities have been observed toward the clumps linked with the shell-like structure and 
the elongated filament ``nw-fl".
\section{Discussion}
\label{sec:disc}
Several authors have reported that the infrared clusters are found at the junction of filaments or filament mergers 
in the star-forming complexes, such as Taurus, Ophiuchus, Rosette, SDC335.579-0.292, W42, Sh 2-138, and IRAS 05480+2545 etc. \citep[e.g.][]{myers09,schneider12,peretto13,dewangan15,baug15,dewangan17b}.  
This particular result favours the role of filaments in the formation of star-forming clumps and clusters of YSOs. 
In the I05463+2652 site, using the {\it Herschel} sub-mm images, the embedded shell-like morphology and filamentary features (including ``nw-fl" and ``s-fl") are revealed.
By estimating the virial mass of the cloud with the help of the previously published the NANTEN $^{13}$CO (beam size $\sim$2$'$.7) derived results, we find the molecular cloud associated with 
I05463+2652 is unstable against gravitational collapse (see Section~\ref{subsec:u1}). 
The embedded shell-like structure contains several massive clumps and many of these massive clumps are 
linked with the areas of high column density in the shell-like structure (see Section~\ref{subsec:temp}), 
where star formation activities are also evident (see Section~\ref{subsec:phot1}).
Furthermore, a spatial association of filament(s) with the areas of high column density has also been found (see Section~\ref{subsec:temp}). 
Note that we have used deep UKIDSS GPS NIR data to trace the embedded YSOs, however, the I05463+2652 site does 
not contain any rich clusters (having peak value of surface density $>$ 100 YSOs/pc$^{2}$). 
At least four groups/clusters of YSOs are seen toward the shell-like structure, indicating that the formation and early evolution of 
young stellar cluster is being occurred in this site. 
Therefore, to infer initial conditions of cluster formation, it will be promising to explore this site using high-resolution molecular line data. Such observations can be helpful to probe the kinematics and 
velocity field of the molecular gas within the molecular cloud linked with the I05463+2652 site. 

In recent years, there has been an increasing interest to identify embedded filaments at an early stage of fragmentation, 
where star formation activities are not yet started. \citet{kainulainen16} studied Musca molecular cloud using the NIR, 870 $\mu$m 
dust continuum and molecular lines data, and suggested that the Musca cloud is a very promising candidate for a filament 
at an early stage of fragmentation and shows very little sign of ongoing star formation.
However, the identification of such filaments is still limited in the literature \citep[e.g.,][]{kainulainen16}. 
In the I05463+2652 site, we have selected two elongated filamentary features (``s-fl" and ``nw-fl").
The {\it Herschel} temperature map reveals these filaments with a temperature of $\sim$11~K.
It has been suggested that the thermally supercritical filaments (i.e., $M_{\rm line} > M_{\rm line,crit}$) can be unstable to radial collapse and fragmentation \citep{andre10}. 
However, thermally subcritical filaments (i.e., $M_{\rm line} < M_{\rm line,crit}$) may lack prestellar clumps/cores and embedded protostars \citep{andre10}. 
In the I05463+2652 site, the observed masses per unit length of the filaments are computed to be 
$\sim$60 M$_{\odot}$ pc$^{-1}$ (for filament ``s-fl") and $\sim$200 M$_{\odot}$ pc$^{-1}$ (for filament ``nw-fl"), 
which are much higher than the critical mass per unit length at T = 10 K (i.e. M$_{\rm line,crit}$ = 16 M$_{\odot}$ pc$^{-1}$) (see Section~\ref{subsec:temp}). 
For a comparison purpose, we also provide the observed masses per unit length of some well known filaments
(such as DR21 ($M_{\rm line}$ $\sim$4500 M$_{\odot}$ pc$^{-1}$), Serpens South ($M_{\rm line}$ $\sim$290 M$_{\odot}$ pc$^{-1}$), Taurus B211/B213 ($M_{\rm line}$ $\sim$50 M$_{\odot}$ pc$^{-1}$), and Musca ($M_{\rm line}$ $\sim$20 M$_{\odot}$ pc$^{-1}$)) \citep[see Table~1 in][]{andre16}. 
Based on the analysis of masses per unit length, the elongated filaments (``s-fl" and ``nw-fl") are thermally supercritical (see Section~\ref{subsec:temp}).
Furthermore, these two filaments contain clumps (or fragments), indicating the signature of gravitational fragmentation. \citet{arzoumanian13} pointed out that supercritical filaments may undergo gravitational contraction and increase in mass per unit length through accretion of background material.

Additionally, the noticeable ongoing star formation is seen along the filament ``nw-fl", while 
there is no signature of star formation found toward the filament ``s-fl" (see Section~\ref{subsec:srf}), 
indicating that these two filaments represent two different evolutionary stages. 
Hence, the lack of star formation toward the clumps (i.e. starless clumps) in filament ``s-fl" makes it a candidate at an early stage of fragmentation. 

To further examine our indicative results, a thorough investigation of the I05463+2652 site will be helpful using high-resolution molecular line observations.
\section{Summary and Conclusions}
\label{sec:conc}
In the present work, we have studied embedded filaments, clumps, and YSOs in the I05463+2652 site as well as 
the surrounding physical environment, using the multi-wavelength data. 
In particular concerning the I05463+2652 site, this work allows for the first time to probe the ongoing physical processes. 
The major results of our multi-wavelength analysis are the following:\\
$\bullet$ A shell-like structure around I05463+2652 is observed in the {\it Herschel} column density (N(H$_{2}$)) map, which is not linked with any ionized emission. \\
$\bullet$ Several noticeable parsec-scale filaments are detected in the {\it Herschel} image at 250 $\mu$m.
At least four areas of high column density ($\gtrsim$ 5 $\times$ 10$^{21}$ cm$^{-2}$; corresponding A$_{V}$ $\sim$5.3 mag) in the shell-like structure are found, where a spatial association with filament(s) is also observed.\\
$\bullet$ Additionally, two elongated filamentary features (``s-fl" (length $\sim$6.4 pc) and ``nw-fl" (length $\sim$8.8 pc)) 
are also identified in the {\it Herschel} sub-mm images 
and are seen more spatially isolated from the shell-like feature. 
The {\it Herschel} temperature map traces the embedded filaments and shell-like structure in a temperature range of about 11--13 K.\\
$\bullet$ 39 clumps are found in our selected site around I05463+2652 and their masses vary 
between 70 and 945 M$_{\odot}$. A majority of clumps (having M$_{clump}$ $\gtrsim$ 300 M$_{\odot}$) 
are spatially distributed toward the shell-like feature.\\
$\bullet$ Both the elongated filamentary features are thermally supercritical. 
The observed masses per unit length of the filaments are computed to be 
$\sim$60 M$_{\odot}$ pc$^{-1}$ (for filament ``s-fl" ) and $\sim$250 M$_{\odot}$ pc$^{-1}$ (for filament ``nw-fl" ), 
which are much higher than the critical mass per unit length at T = 10 K (i.e. M$_{\rm line,crit}$ = 16 M$_{\odot}$ pc$^{-1}$).\\
$\bullet$ The elongated filament ``nw-fl" is fragmented into five clumps that have masses between $\sim$100--545 M$_{\odot}$, 
while the other filament ``s-fl" is fragmented into two clumps that have masses of $\sim$170 and 215 M$_{\odot}$.\\
$\bullet$ Based on the analysis of photometric 1--5 $\mu$m data, 175 YSOs are identified in the I05463+2652 site. 
Star formation activities are evident toward the four areas of high column density in the shell-like structure, where massive clumps are present.\\
$\bullet$ The clumps seen toward the elongated filament ``nw-fl" are associated with YSOs, 
while the clumps observed toward the filament ``s-fl" do not show any signature of star formation activities (i.e. starless clumps).

Taken together, our observed results reveal the role of filaments in star formation process in the I05480+2545 site. 
Furthermore, we have also investigated a promising filament candidate (i.e. filament ``s-fl") at an early stage of fragmentation. 
\acknowledgments 
We thank the anonymous reviewer for constructive comments and suggestions. 
The research work at Physical Research Laboratory is funded by the Department of Space, Government of India. 
This work is based on data obtained as part of the UKIRT Infrared Deep Sky Survey. This publication 
made use of data products from the Two Micron All Sky Survey (a joint project of the University of Massachusetts and 
the Infrared Processing and Analysis Center / California Institute of Technology, funded by NASA and NSF), archival 
data obtained with the {\it Spitzer} Space Telescope (operated by the Jet Propulsion Laboratory, California Institute 
of Technology under a contract with NASA). {\it WISE} is a joint project of the
University of California and the JPL, Caltech, funded by the NASA. {\it Herschel} is an ESA space observatory with science instruments provided by European-led Principal Investigator consortia and with important participation from NASA. RD acknowledges CONACyT(M\'{e}xico) for the PhD grant 370405. 
%
%


\begin{deluxetable}{ccccc}
\tablewidth{0pt} 
\tablecaption{Physical parameters of the identified {\it Herschel} clumps in our selected field around I05463+2652 (see Figure~\ref{fig3x}d). 
Column~1 gives the IDs assigned to the clump. Table also provides 
Galactic coordinates (l, b), deconvolved effective radius (R$_{clump}$), and clump mass (M$_{clump}$). 
The clumps associated with the elongated filament ``nw-fl" are highlighted with dagger 
symbols (see ID nos 2, 4, 6, 8, and 9, and also Figure~\ref{fig3x}d), while the clumps linked with the elongated filament ``s-fl" 
are marked by double-dagger symbols (see ID nos 35 and 36, and also Figure~\ref{fig3x}d).\label{tab1}} 
\tablehead{ \colhead{ID} & \colhead{{\it l}} & \colhead{{\it b}} & \colhead{R$_{c}$}& \colhead{M$_{clump}$}\\
\colhead{} &  \colhead{[degree]} & \colhead{[degree]} & \colhead{(pc)} &\colhead{($M_\odot$)}}
\startdata 
          1    &   182.077   	 &   -0.093	    &    0.7	    &    75         \\
          2$\dagger$    &   182.038   	 &   -0.113	    &    0.8	    &    100       \\
          3    &   182.088   	 &   -0.140	    &    0.9	    &   145       \\
          4$\dagger$    &   182.034   	 &   -0.152	    &    1.2	    &   335       \\
          5    &   182.069   	 &   -0.175	    &    1.0	    &   195       \\
          6$\dagger$    &   182.030   	 &   -0.210	    &    1.5	    &   545       \\
          7  	&  181.976   	  &  -0.264	     &   0.8	     &   85       \\
          8$\dagger$    &   182.053   	 &   -0.292	    &    1.3	    &   485       \\
          9$\dagger$    &   182.065   	 &   -0.338	    &    1.0	    &   295       \\
         10    &   182.085   	 &   -0.358	    &    1.4	    &   450       \\
         11    &   182.030   	 &   -0.354	    &    1.0	    &   145       \\
         12  	&  181.976   	  &  -0.385	     &   1.0	     &  155       \\
         13    &   182.100   	 &   -0.428	    &    1.5	    &   345       \\
         14    &   182.077   	 &   -0.490	    &    1.0	    &   195       \\
         15  	&  182.065   	  &  -0.517	     &   0.6	     &   70       \\
         16  	&  181.999   	  &  -0.494	     &   0.7	     &   90       \\
         17    &   182.034   	 &   -0.529	    &    1.0	    &   180       \\
         18    &   182.018   	 &   -0.583	    &    1.3	    &   370       \\
         19    &   182.069   	 &   -0.541	    &    0.8	    &   120       \\
         20    &   182.088   	 &   -0.583	    &    1.1	    &   215       \\
         21    &   182.127   	 &   -0.552	    &    1.4	    &   330       \\
         22    &   182.174   	 &   -0.471	    &    1.7	    &   670       \\
         23  	&  182.263   	  &  -0.412	     &   1.9	     &  710       \\
         24  	&  182.213   	  &  -0.385	     &   1.3	     &  370       \\
         25    &   182.174   	 &   -0.373	    &    1.2	    &   330       \\
         26    &   182.182   	 &   -0.346	    &    1.1	    &   290       \\
         27    &   182.201   	 &   -0.327	    &    1.2	    &   315       \\
         28    &   182.244   	 &   -0.276	    &    1.4	    &   390       \\
         29    &   182.260   	 &   -0.292	    &    1.1	    &   250       \\
         30    &   182.291   	 &   -0.319	    &    1.9	    &   945       \\
         31    &   182.357   	 &   -0.385	    &    1.1	    &   195       \\
         32    &   182.326   	 &   -0.412	    &    1.2	    &   210       \\
         33    &   182.423   	 &   -0.459	    &    1.0	    &   160       \\
         34    &   182.314   	 &   -0.509	    &    1.2	    &   210       \\
         35$\ddagger$    &     182.170 	 &     -0.630	    &      1.1      &     215       \\
         36$\ddagger$    &     182.170 	 &     -0.751	    &      1.0      &     170       \\
         37    &     182.244 	 &     -0.712	    &      0.9      &     125       \\
         38    &     182.213 	 &     -0.801	    &      0.9      &     130       \\
         39    &     182.030 	 &     -0.782	    &      0.7      &      75       \\
 \enddata  
\end{deluxetable}


\begin{thebibliography}{}
%
\bibitem[Andr{\'e} et al.(2010)]{andre10}
Andr{\'e} P., Men'shchikov A., Bontemps S., et al., 2010, A\&A, 518, L102

\bibitem[Andr{\'e} et al.(2014)]{andre14}
Andr{\'e}, P., Di Francesco, J., Ward-Thompson, D., et al. 2014, in Protostars and Planets VI, ed. H. Beuther et al. (Tucson, AZ; Univ. Arizona Press), 27

\bibitem[Andr{\'e} et al.(2016)]{andre16}
Andr{\'e} P., Rev{\'e}ret V., K\"{o}nyves V., et al., 2016, A\&A, 592, 54

\bibitem[Arzoumanian et al.(2013)]{arzoumanian13}
Arzoumanian D., Andr{\'e} P., Peretto N., K\"{o}nyves V., 2013, A\&A, 553, 119

\bibitem[Assirati et al.(2014)]{assirati14}
Assirati, L., Silva, N.~R., Berton, L., Lopes, A.~A., \& Bruno, O.~M. 2014, Journal of Physics: Conference Series, 490(1), 2014

\bibitem[Baug et al.(2015)]{baug15}
Baug T., Ojha D.~K., Dewangan L.~K., et al., 2015, MNRAS, 454, 4335

\bibitem[Bohlin et al.(1978)]{bohlin78}
Bohlin R.~C., Savage B.~D., Drake J.~F., 1978, ApJ, 224, 13233

\bibitem[Bressert et al.(2010)]{bressert10}
Bressert E., Bastian N., Gutermuth R., et al., 2010, MNRAS, 409, 54

\bibitem[Condon et al.(1998)]{condon98}
Condon J.~J., Cotton W.~D., Greisen E.~W., et al., 1998, AJ, 115, 1693

\bibitem[Contreras et al.(2016)]{contreras16}
Contreras Y., Garay G., Rathborne J.~M., Sanhueza P., 2016, MNRAS, 456, 2041

\bibitem[de Graauw et al.(2010)]{degraauw10}
de Graauw T., Helmich F.~P., Phillips T.~G., et al. 2010, A\&A, 518, L4

\bibitem[Dewangan et al.(2015)]{dewangan15}
Dewangan L.~K., Luna A., Ojha D.~K., et al., 2015, ApJ, 811, 79

\bibitem[Dewangan et al.(2017a)]{dewangan17a}
Dewangan L.~K., Ojha D.~K., Zinchenko I., Janardhan P., Luna A.,  2017a, ApJ, 834, 22

\bibitem[Dewangan et al.(2017b)]{dewangan17b}
Dewangan L.~K., Ojha D.~K., \& Baug, T.,  2017b, ApJ, 844, 15

\bibitem[Flaherty et al.(2007)]{flaherty07}
Flaherty K.~M., Pipher J.~L., Megeath S.~T., et al., 2007, ApJ, 663, 1069

\bibitem[Gonzalez \& Woods(2011)]{gonzalez11}
Gonzalez, R, \& Woods, R. 2011, {\it Digital Image Processing} (Pearson Education) ISBN 9780133002324

\bibitem[Griffin et al.(2010)]{griffin10} 
Griffin M.~J., Abergel A., Abreu A, et al., 2010, A\&A, 518, L3

\bibitem[Gutermuth et al.(2009)]{gutermuth09}
Gutermuth R.~A., Megeath S.~T., Myers P.~C., et al., 2009, ApJS, 184, 18

\bibitem[Hildebrand(1983)]{hildebrand83} 
Hildebrand R.~H., 1983, QJRAS, 24, 267

\bibitem[Inutsuka \& Miyama(1997)]{inutsuka97}
Inutsuka S., Miyama S.~M., 1997, ApJ, 480, 681

\bibitem[Kainulainen et al.(2016)]{kainulainen16}
Kainulainen J., Hacar A.,  Alves J., et al., 2016, A\&A, 586, 27

\bibitem[Kainulainen et al.(2017)]{kainulainen17}
Kainulainen J., Stutz, A.~M., Stanke, T., et al., 2017, A\&A, 600, 141

\bibitem[Kauffmann et al.(2008)]{kauffmann08}
Kauffmann J., Bertoldi F., Bourke T.~L., Evans II, N.~J., Lee, C.~W., 2008, ApJ, 487, 993

\bibitem[Kawamura et al.(1998)]{kawamura98}
Kawamura A., Onishi T., Yonekura Y. et al., 1998, ApJS, 117, 387

\bibitem[Lawrence et al.(2007)]{lawrence07}
Lawrence A., Warren S.~J., Almaini O., et al., 2007, MNRAS, 379, 1599

\bibitem[Li et al.(2016)]{li16}
Li Guang-Xing, Urquhart J.~S., Leurini S., et al., 2016, A\&A, 591, 5

\bibitem[MacLaren et al.(1988)]{maclaren88}
MacLaren I., Richardson K.~M., Wolfendale A.~W., 1988, ApJ, 333, 821

\bibitem[Mallick et al.(2015)]{mallick15}
Mallick K.~K., Ojha D.~K., Tamura M., et al., 2015, MNRAS, 447, 2307

\bibitem[Molinari et al.(2010)]{molinari10}
Molinari S., Swinyard B., Bally J., et al., 2010, A\&A, 518, L100

\bibitem[Myers (2009)]{myers09} 
Myers P.~C., 2009, ApJ, 700, 1609

\bibitem[Ostriker(1964)]{ostriker64}
Ostriker J., 1964, ApJ, 140, 1056

\bibitem[Ott(2010)]{ott10}
Ott S., 2010, in Astronomical Society of the Pacic Conference
Series, Vol. 434, Astronomical Data Analysis Software and
Systems XIX, ed. Y. Mizumoto, K.-I. Morita, \& M. Ohishi, 139

\bibitem[Peretto et al.(2013)]{peretto13}	
Peretto N., Fuller G.~A., Duarte-Cabral A., et al., 2013, A\&A, 555, 112

\bibitem[Pilbratt et al.(2010)]{pilbratt10}	
Pilbratt G.~L., Riedinger, J.~R., Passvogel, T., et al., 2010, A\&A, 518, L1

\bibitem[Poglitsch et al.(2010)]{poglitsch10}	
Poglitsch A., Waelkens C., Geis N., et al., 2010, A\&A, 518, L2

\bibitem[Ragan et al.(2014)]{ragan14}
Ragan S.~E., Henning Th., Tackenberg J., et al., 2014, A\&A, 568, 73

\bibitem[Schneider et al.(2012)]{schneider12}
Schneider N., Csengeri T., Hennemann M., et al., 2012, A\&A, 540, L11

\bibitem[Skrutskie et al.(2006)]{skrutskie06}
Skrutskie M.~F., Cutri R.~M., Stiening R., et al., 2006, AJ, 131, 1163

\bibitem[Whitney et al.(2011)]{whitney11}
Whitney B., Benjamin R., Meade M., et al., 2011, BAAS, 43, 241.16

\bibitem[Williams et al.(1994)]{williams94} 
Williams J. P., de Geus E. J., Blitz L., 1994, ApJ, 428, 693

\bibitem[Wright et al.(2010)]{wright10}
Wright E.~L., Eisenhardt P.~R.~M., Mainzer A.~K., et al., 2010, AJ, 140, 1868
%
\end{thebibliography}
 \end{document}